# A method to constrain the total mass of galaxy groups

Giovanni C. Baiesi Pillastrini[1]
*Sezione Spettroscopia Astronomica - U.A.I.  c/o Dip. Astronomia, Vicolo Osservatorio, 5   35122 Padova Italy*
email: gcbp@it.packardbell.org

[1] permanent address: via Garzoni, 2/2  40138 Bologna  Italy

Abstract

We present a simple method to constrain the total mass of groups of galaxies. Tidal theory predicts that a limit to the mass of bound groups of galaxies can be obtained by using the fact that the tidal forces due to the external mass distributions are insufficient to disrupt the groups. To illustrate how the method works, we find tidal limits on the mass of eleven nearby galaxy groups. In most cases, tidal limits placed on these groups show that mass estimations obtained from methods based either on the application of the virial theorem or moments of the projected mass are underestimated by a factor of ~2 even if, in many cases, errors are large. Three groups show virial parameters fully concordant with the tidal constraints while two outlier groups show anomalous results. The "irregular" state of the latter suggests that the reliability of the method depends on the physical properties of the test groups, which should match the fundamental assumptions of spherical symmetry and dynamical equilibrium state.

1. Introduction

Since the pioneering works by Zwicky (1933) and Smith (1935), mass estimations of galaxy groups and clusters are generally determined by methods based on the application of the virial theorem to the internal dynamics (see also Jackson, 1975; Rood & Dickel, 1978). In order to reduce the statistical uncertainties affecting virial estimators, Bahcall & Tremaine (1981) and Heisler et al. (1985) suggested alternative methods based on the moments of the projected mass. It is well known that all these methods suffer of large uncertainties mainly due to the difficulty of inferring the group memberships using kinematic data alone. Indeed, Cen (1997) and Diaferio et al. (1999), utilizing large scale simulations to investigate the projection effects on various observable parameters (harmonic radius, velocity dispersion, mass ect.) of groups and clusters found that projection alter them in different ways to varying extents. But, as expected, projection effects increase at larger radii where the contamination due to interlopers becomes higher. In particular, for galaxy groups, Cen found that this bias induces a substantial overestimation of projected sizes, an underestimation of velocity dispersions and a resulting 25% underestimation of their masses.

Similarly, the $R_o$-method based on the spherical infall model (Lynden-Bell 1981, Sandage 1986) needs a careful determination of the turnaround radius of the system. Still, interlopers and selection effects largely affect this mass estimator technique that strictly depends on the identification of the true group members. In other words, to reach a high accuracy, a complete velocity-distance relation of the galaxies surrounding the main body of the group should be necessary for memberships confirmation. This accuracy was fairly obtained to estimate the mass of the Local Group (Karachentsev & Kashibadze, 2005) and nearby groups of galaxies (Karachentsev, 2005), but hardly for groups farther away. This is a very difficult task when one studies systems at intermediate distances for which mass measurements turn out systematically underestimated (Frenk et al. 1996).

Hartwick (1976, 1981) suggested an alternative strategy based on the tidal relationships between two gravitationally interacting bodies, where mass of clusters may be established by examining the tidal limits set by the clusters on nearby groups of galaxies. He derived the upper limit to the mass of Virgo and Coma clusters by considering the absence of breaking tidal effects of the clusters on nearby bound groups. The accuracy of this method depends directly on the reliability of the virial mass estimated for the test groups, which is the topic of the present study.

In line with the above, we attempt a similar approach knowing that a self-gravitating system is dependent on a proper representation of the external gravitational influence. We follow the strategy involving approximate descriptions of external influences incorporating the larger external influence through static, external tidal field estimated on the basis of the present-day locations of an extended sample of galaxies representative for the external density distribution. We examine the general expression of the tidal force generated by the galaxy distribution in an arbitrary spherical sampling volume.

For a sphere centered on a galaxy group, numerical calculations and simulation technique will attempt to overcome volume incompleteness by computing the amplitude of the tidal force within a series of concentric spheres. What is measured is the cumulative amplitude of the tidal force, which is expected to converge asymptotically within the boundary of the finite sampling volume. From an observational point of view, tidal radii imposed by external tidal fields to the extensions of virialized groups must be *greater or, at least comparable* with radii defined by direct observation of the group member positions. On the contrary, in the absence of tidal disruption effects, from the tidal relationships between internal and external gravitational forces we can obtain direct information about the total mass of the group.

To disentangle this issue, we present and discuss inferences drawn regarding the total mass determination of galaxy groups. The method is described in Ch.2. An application of the method to a sample of eleven nearby groups is performed in Ch.3. The results are discussed in Ch.4 and the conclusions are drawn in Ch.5.

2. The method

The present model assumes that the source of the tidal force is due to a time-independent host gravitational potential generated by the surrounding external matter distribution enclosed within a fixed spherical volume centered on a galaxy group. Of course, a small system of galaxies as a group does not suffer any kind of tidal effect if the distribution of inhomogeneities outside the group has an effective spherical symmetry. However, this symmetry holds only on average, so that, in principle, there could be a non-vanishing effect on the group due to deviation from perfect isotropy. Our goal, therefore, is to study the influence of such anisotropies on the groups and, in particular, whether they give rise to tidal limits comparable with the observed extensions of the groups. Generally, assuming the tidal effect as a static tidal limitation, the tidal radius is defined as the distance from the center of a "satellite" (here, a galaxy group assumed to move in a straight line orbit toward or away the center of mass of the host galaxy distribution) beyond which the tidal effect of the host potential exceed the self-gravity of the satellite. Therefore, the fundamental quantity now is the gravitational potential created by the distant inhomogeneities.

At this point, we introduce an important simplification: rather than carrying out a fully relativistic analysis, we will work in the Newtonian approximation in the reference frame of the groups. This is justified by the assumption that the involved velocities are non relativistic. Then, the tidal force acting on a galaxy group can be expressed by

$$F_{tidal,a} \equiv -\frac{d^2\Phi_{ext}}{dR_a dR_b} R_b \equiv F_{ab} R_b \qquad (1)$$

where $\Phi_{ext}$ is the external potential and $R$ is the radius vector in the group reference frame. Now, considering a group at position vector $r$ from the observer under the action of $N$ surrounding galaxies at a position $r_g$ and mass $m_g$, the external tidal potential is given by

$$\Phi_{ext} = -G \sum_N \frac{m_g}{|r_g - r|} \qquad (2)$$

and, the tidal tensor is

$$F_{ab} = \sum_N \left( \frac{m_g}{|r_g - r|^3} \right) \left( \frac{3(r_g - r)_a (r_g - r)_b}{|r_g - r|^2} - \delta_{ab} \right) \qquad (3)$$

where the gravitational constant $G = 1$ and $\delta_{ab}$ is the Kronecker delta. It follows that the amplitude of the tidal force is given by

$$F_{tidal} = |F_{aa} R_a| \qquad (4)$$

where $F_{aa}$ are the three eigenvalues corresponding to the principal axes of the 3 x 3 symmetric matrix $F_{ab}$. Summation convention (for which any product of terms a repeated suffix is held to be summed over its three values 1, 2 and 3 and a suffix not repeated can take any of those values) is assumed for Eq.(1) and (4).

By assuming that a test group should be "regular" i.e. spherically symmetric and dynamically relaxed, the condition $F_{tidal} = F_{binding}$ must be satisfied. Plugging in

$$F_{tidal} = \frac{M}{R^2} \qquad (5)$$

Then, the tidal radius is

$$R_t = \left( \frac{M}{F_{tidal}} \right)^{\frac{1}{2}} \qquad (6)$$

where $M$ and $R$ are the mass and radius of the test group.

Note that, paradoxically, the robustness of the method is established by the uncertainties affecting the evaluation of the external tidal forces. That is, a significant underestimation of $F_{tidal}$ can be due to

several bias effects as:
i) an insufficient extension of the sampled spheres, which may prevent its convergence;
ii) selection and incompleteness effects which flatten its estimate especially for distant groups; iii) inappropriate mass weighting of the galaxy distribution within the sampled spheres which is responsible for the major share of the tide (we discuss it later). In other words, we have a high probability to obtain, if not the true, the l*ower* $F_{tidal}$ limit as well as (consequently) the *upper* $R_t$ limit! Therefore, two cases can be evaluated:

i) $R/R_t \leq 1$, then, the internal binding force balances the external tidal perturbation. It follows that the group is expected to be in a stable and relaxed configuration and its estimated mass $M$ is appropriate (even if an underestimated value of $F_{tidal}$ can give the same result as pointed out later);

ii) $R/R_t > 1$, then, the configuration of the group is expected to show signs of tidal stripping and disruption. Of course, this is not the case by assumption. The inconsistency between the observed and predicted radii could be explained either by a non-relaxed state of the group (in which case, the group should be ruled out from analysis) or by incomplete kinematic information of the group memberships (in which case, $R$ and $M$ turn out underestimated). If the latter is the case, the "gap" should be evaluated in term of *mass deficiency*. That is, assuming $F_{tidal}$ as the *true* tidal effect, if the radius $R$, now assumed as the *lower "observable" tidal radius limit* takes place of the *upper "predicted" tidal radius* $R_t$ in Eq. (6) and solving for $M$ then, we will obtain a new mass estimation, say $M_{total}$, which is expected to be the *"lower" total mass* of the groups. Therefore, when $R/R_t > 1$, we will detect the "missing" mass fraction simply by the parameter $\mu = M_{total}/M$ or, plugging in, $(R/R_t)^2$. The evidence that we are really constraining the lower total mass of a group is supported by the fact that either $R/R_t$ and $\mu$ are (lower) limited parameters.

According to this conservative line of analysis, we neglect the contribution to the tidal forces due to a non-zero Lambda cosmology.

## 3. Application

### 3.1. The data

In order to give a simplified illustration of how the method may work, firstly we select a small but fair sample of galaxy groups from the UZC-SSRS2 group catalog (Ramella et al. 2002). The catalog contains 1168 groups assembled using the friends of friends algorithm with a linking parameter that scales with increasing redshift in order to take into account the galaxy selection function. Secondly, we need to map the galaxy distribution that extends well beyond the region of the selected groups. Datasets are extracted from the catalog of published redshift measurements of galaxies collected by Huchra et al. (1995, ZCAT hereafter). This catalog is magnitude-limited in the parts filled by the CFA2 redshift survey (Geller & Huchra, 1989). Therefore, we limit the regions of selection within the boundaries of this survey, which is complete up to $B \leq 15.5$. We assess a local spherically shaped 20 $h^{-1}Mpc$ radius ("sphere" hereafter) centered on a test group within which the redshift survey is expected to encapsulate the major share of matter inhomogeneities responsible for the external tidal forces. We adopt the following criteria to select the test groups.

i) They should lie at $z \leq 0.015$ under which the bias due to the galaxy selection function would not change significantly our results. This bias is introduced by the magnitude limit of the catalog in which the spheres of fixed linear radius contain different counts of observed galaxies at different redshift so that, at large radii, volume incompleteness may flatten the amplitude of the external tidal perturbation.
ii) The groups should have at least 8 or more member galaxies to avoid large errors in the mass estimations due to small number statistics.
We have found eleven groups that match these criteria: UZC268, 291, 323, 331, 340, 376, 466, and 478, 480, 518, and 578. These groups lie in the northern hemisphere within the boundaries defined by the CFA2 survey. Virgo and Ursa Major clusters are the principal mass concentrations responsible for the major share of the external gravitational influences in this region.

3.2. Simplifying assumptions

For practical reasons, since we are merely interested in a simplified application of the method, the following simplifying assumptions have been adopted:
i) Exact evaluations of tidal forces require a *complete* and *real* spatial distribution of galaxies. Unfortunately, the presence of large peculiar velocity fields in the local Universe induces the so-called "redshift distortion" effect which largely bias any distribution of galaxies generated from redshift catalogs. There are many methods to reconstruct an approximated real galaxy distribution (see for instance Dekel et al.1990, Nusser & Branchini 2000, Peebles 1989, Willick et al. 1997 among others). However, accounting for large datasets would become prohibitively expensive in terms of computational effort for conventional reconstructions. Then, we restrict ourselves to a case study example limiting our effort in a conventional linear analysis.
ii) We assume that the local velocity field takes the form of a collapsing streaming motion towards the Virgo cluster so that each radial velocity can be easily corrected for this effect (on the other hand, if we assume the peculiar velocity field as independent from direction, it will average to zero within the spheres). It is common to work in the rest frame implied by the microwave dipole. Here we work in a local rest frame. Since our objects are so local, the dominant motion would be a common flow in the microwave background rest frame and the difference would be only a coordinate translation. Therefore, distances are calculated by the conventional $cz/(100h)$ where $cz$ is the heliocentric radial velocity corrected for the
Virgocentric infall (Terry et al., 2002) and $h$ is the Hubble constant in unit of $100 Kms^{-1}Mpc^{-1}$. iii) We limit our investigation in the local Universe where the galaxy population could be assumed to trace the underlying density field (Hoekstra et al., 2001, Verde et al., 2002). This assumption holds for statistical purpose so that mass is approximately concentrated in $N$ discrete points defined by the spatial distribution of galaxies within the spheres beyond which the Universe doesn't have significant gravitational influence on the groups.
iv) To avoid an underdensity in the zone of obscuration due to lost information, the masked regions at low Galactic latitudes are filled in with a random distribution of synthetic galaxies having the observed mean number density.
v) In Eq. (3) we assume that the local density perturbation is approximated by the summation of individual galaxy masses. This assumption could be objected since structures like groups and clusters of galaxies are present inside the spheres (Virgo, Ursa Major, NGC 5371 group etc.). Generally, a significant mass component due to hot intracluster gas is present within these systems and could have a relevant influence in the determination of the total mass (Helsdon et al. 2005). Conventionally, the mass of these structures should be evaluated dynamically as a whole although the error might be even larger than that arising from the summation of individual contributions. In a clumpy distribution of galaxies, as the local universe is, this procedure should add more weight to the external mass density increasing the strength of the tidal perturbation and, in turn, the amplitude of $F_{tidal}$ (and its uncertainty). Then, we expect that the amplitude of the tidal force predicted by the summation of individual masses would be underestimated favoring our conservative approach for which the analysis should have more significance if the estimate of the tidal force will be represented by its lower limit.

### 3.3. Error in estimating the tidal force

The galaxy mass $m_g$ is the only free parameter in Eq. (3). For statistical purpose, galaxy mass could be assumed proportional to its luminosity so that the contribution may be weighted by its luminosity. However, luminosity weighting has little physical justification since luminosity is not well correlated with mass and increases the statistical error (Bahcall & Tremaine, 1981). In alternative, one may assign a common averaged mass to the sample of galaxies under analysis (Raychaudhury & Lynden-Bell, 1989), but this approach cannot give information about the uncertainty affecting the evaluation of the tidal force from an unknown mass distribution.

Use of simulation techniques can help us to estimate statistically the error introduced by $m_g$ in the calculation of Eq.(3). To assess the variability of the tidal force we proceed as follows: for each test group we simulate $n = 1000$ sampled spheres assigning to the $N$ galaxies within each sphere a random values of $m_g$ picked from an uniformly distributed interval $[2 \cdot 10^{11}, 3 \cdot 10^{12}] \cdot h^{-1} M_{sun}$ while their 3D-positions remains fixed. Then, $F_{tidal}$ and its standard deviation $\sigma_{tidal}$ will be computed averaging the $n$ outputs of Eq. (4) obtained from the $n$ synthetic samples so constructed. The mass interval is assumed to approximate the real range of galaxy mass from late to early morphological types excluding extreme values for dwarf and giant objects.

### 3.4. Results

Results are summarized in Table 1. From col. (1) to col. (7) it list the parameters of the UZC groups used in the application: Group identification, number of galaxies within groups ($N_g$), equatorial coordinates J2000), heliocentric radial velocity (V), velocity dispersion ($\sigma_V$), virial radius ($R_v$) obtained from $2 R_h N_g / (N_g - 1)$ where $R_h$ is the harmonic radius and, the virial mass ($M_v$) obtained from $3\sigma^2 R_v / G$. These data are taken from Table 1 of the UZC-SSRS2 group catalog (Ramella et al., 2002). In col. (8) the amplitude of the tidal force $F_{tidal}$. The errors represent the $1\sigma_{tidal}$ scatter around the average value as defined in 3.3. In col. (9) the tidal radius $R_t$ and its error. In col. (10) the ratio $R_v / R_t$ and, in col. (11) the mass fraction $\mu = M_{total} / M_v = (R_v / R_t)^2$ and its $1\sigma$ error. The error of $\mu$ as well as the tidal radius error is estimated using error propagation technique assuming a standard fractional error of $\pi^{-1}(2 \ln N_g / N_g)^{1/2}$ for $M_v$ (Bahcall & Tremaine, 1981) and an arbitrary 20% internal error on the estimation of the virial radius $R_v$.

### 3.5. the convergence of the tidal force

It is crucial to see to what extent the depicted galaxy distribution can indeed be held responsible for most of the inferred tidal perturbation. To test it, the tidal forces are computed for a set of concentric external spheres of increasing $2h^{-1}Mpc$ radii from the center of the group. Fig.1 shows the developments of their cumulative amplitudes as a function of external distances. Note that they are systematically convergent within ~10-15 $h^{-1}Mpc$ radii and the spheres are large enough to incorporate the major share of the gravitational influence exerted by the inhomogeneous matter distribution. This is in line with our earlier assumption that there was hardly noticeable tidal contribution from large distances.

The large variation of the tidal amplitudes can be explained simply by the fact that the nearer a test group will be to large mass concentrations, the higher the amplitude of the external tidal force will be. One could object that some bias effects may hide the true convergence so that, in reality, the asymptote would be approached farther away. Also in this case, the tidal force would result underestimated justifying the above assumption of $F_{tidal}$ as the lower limit where the true tidal perturbation converges to its final value. Then, the absence of disruptive tidal effects on the groups having $\mu > 1$ demonstrates a systematic trend in underestimating the virial mass measurements.

3.6. Outliers

Different the case of UZC 268 and 340, which have anomalous large values of $\mu$ (18 and 10, respectively). The groups have $\sigma_V$ = 57 and 81 $Kms^{-1}$ comparable with those noticed in very compact groups of galaxies. These low velocity dispersions can be explained with the so-called "caustics" effect appearing as a by-product of collapsing groups in a pre-virialized state, in which large peculiar velocities of infalling galaxies towards the center provide a "squeezed" representation of the corresponding radial velocities and low amplitudes of the velocity dispersion. In this case, the groups are not in dynamical equilibrium state. An alternative explanation could be that we are observing "chains" or "sheets" like structures of galaxies that, anyhow, do not match the assumption of spherical symmetry. We conclude that both UZC 268 and 340 should be considered outliers since their physical properties do not match the essential assumptions required from the test.

3.7. Comparison between virial and projected mass estimators

Evans et al. (2003), studying tracer populations used to estimate virial and projected masses for different clustered systems, found that groups of galaxies, in particular, are largely biased from projection effects and fail in tracing mass especially at large radii. On the basis of 10 000 Monte Carlo simulations, they demonstrated that at least 87% of the virial mass estimations for groups of galaxies are below the true mass and the alternative method based on the moments of the projected mass gives less underestimations (53%) suggesting that the latter is more accurate than that based on the virial theorem. We have taken this suggestion into account introducing in our analysis a new set of mass estimations obtained from the projected mass estimator. In the case of isotropic orbits, the projected mass $M_{PM}$ is given by $32/\pi G N_g \sum_i V_i^2 R_{\perp,i}$ (Heisler et al.1985) according to the fact that galaxies in groups have, on average, nearly isotropic orbits (Carlberg et al. 2001; Mahdavi & Geller, 2004). Still, for each group, we calculate a new tidal radius and the corresponding, say, $\mu_{PM}$ to evaluate differences with respect to the previous analysis based on the virial data. Visually, this is immediately reflected in the differences between the values of $\mu$ and $\mu_{PM}$. Fig.2. provides such visual comparison. It shows the different efficiency in detecting the total mass of the groups performed by the two mass estimator techniques. In fact, as predicted by the simulations, half of the values of $\mu_{PM}$ (obtained from $M_{PM}$) are less underestimated than those of $\mu$.

4. Discussion

The above comparison (3.7.) does not change significantly our previous results. The systematic trend in underestimating masses independently from the mass estimators used, need to be discussed in more details:

i) Six of the selected UZC groups have the parameter $\mu > 1$ pointing out that a certain mass fraction of these groups is missed from conventional mass estimators. In spite of the low levels of statistical significance, these results should be considered quite reliable for two reasons: first, they lie at a very short distance from the observer so that effects of volume incompleteness due to the galaxy selection function should be negligible. Second, they are subject to significant gravitational perturbations due to their positions in proximity of the Virgo cluster (or Ursa Major in the case of UZC 480 and 518) so that their tidal amplitudes should be close to the true values. From the above considerations and having in mind that we are finding the lower limits of $\mu$, we may conclude that mass measurements obtained for these groups do not account for the whole mass and significant mass fractions remain undetected.

ii) One could object that the assumption of the virial radius $R_v$ as the lower observable tidal radius limit is not appropriate to constrain the real boundary of a galaxy group. Diaferio et al. (1999) studying projection effects on the observable parameters of galaxy groups made a comparison between groups extracted from a 3D simulated galaxy distribution in real space and the CFA2N groups (Ramella et al. 1997) defined in redshift space. They found that harmonic radii $R_h$ in redshift space are biased high by about a factor of 2 due to interloper contamination. This bias supports the idea that $R_v (\propto R_h)$ is systematically overestimated as well as $\mu$, consequently. This conclusion weakens our results. However, precision would require that, if available, the best parameter for our test should be the "turnaround" radius, which limits the location of the zero velocity surfaces around bound systems. In the absence of accurate peculiar velocity maps of the immediate neighbor around the groups to separate bound from unbound galaxies, constraining the turnaround radius from a small set of kinematic data could be a controversial task. In any case, we know that the turnaround radius, as predicted by the spherical infall model, ranges between 2.9 and 3.4 times the virial radius for $0 \leq \Omega_0 \leq 1$ (Praton & Schneider, 1994). Taking into account this theoretical constraint on $R_v$, even if it turns out overestimated from projected effects, it should not be larger than the predicted turnaround radius. Then, still driven from the conservative approach of lowering our results, the use of $R_v$ in estimating $\mu$ can be thought as a reasonable compromise between theoretical predictions and simulated data.

iii) Another important point is concerned with how representative is the computed value of $F_{tidal}$. As stated before, the tidal force is fairly determined when its cumulative amplitude converges asymptotically within the sampled sphere. However, we are aware of possible bias effects due to volume incompleteness at large radii that may slightly underestimate the amplitude of $F_{tidal}$ as well as the value of $\mu$. This could be the case of UZC 331, 466 and 578 for which $\mu < 1.3$. In fact, they are the more distant groups of the sample where the above-discussed bias effect could be present. As evidence, note in Table 1 the low amplitude of $F_{tidal}$ measured for these groups. Also UZC 376 has $\mu < 1$ but it lies at intermediate distance of the group sample. It shows a substantial equilibrium between internal and external gravitational forces and, as expected from a virialized and relaxed group, its virial and projected mass estimates are comparable as can be seen in Fig.2.

iv) One could object our assertion according to which $F_{tidal}$ would be underestimated. Actually, a suspect

may arise on the adopted mass interval used to extract the random value of $m_g$. That is, if one adopts an opposite viewpoint assuming $M_v$ as the *true* mass, then $m_g$ could alter the amplitude of the tidal forces. In other words, to have $R_t \geq R_v$ on all the examined cases, the mass interval from which $m_g$ should be picked up must be decreased by a factor of ~3 with respect to our original choice. Namely, the galaxy mass interval should be rescaled down as $[2 \cdot 10^{11}, 10^{12}] \cdot h^{-1} M_{sun}$! Such an interval does not represent the current galaxy mass estimations obtained from accurate lensing measurements (Coil et al. 2005, Guzik & SeljaK, 2002; Hoekstra et al., 2005, McKay et al. 2001). Therefore, according to the results of Cen (1996), Frenk et al. (1997) and Evans et al. (2003), our method confirms that conventional mass estimators generally underestimate the true mass of galaxy groups.

This persistent discrepancy between the observed and tidal-predicted mass can be explained only with the assertion that $\mu$ reflects a mass fraction located within the group dark halo so far undetected by virial and projected mass estimators.

5. Conclusions

The main accomplishment of this paper is the introduction of a method based on the tidal theory where the total mass of galaxy groups is tidally constrained by the external mass distributions. This is also a simple and independent way to test the reliability of mass estimations obtained from different and independent methods. The advantage of our method resides in the fact that it gives the lower limit to the amplitude of external tidal field acting on a galaxy group. In fact, such amplitude depends on whether the galaxy distribution in the flux limited galaxy catalog does represent an unbiased reflection of the actual external mass distribution surrounding the groups. In other words, the amplitude of tidal forces may turns out underestimated by the combined effects of mass underweighting and volume incompleteness of the underlying galaxy distribution. Fortunately, such expectation favors a conservative estimation of the tidal force providing an upper limit to the tidal radius and, in turn, a lower limit to the total mass of a galaxy group. We have applied our method to a sample of eleven groups. For six groups, masses evaluated by our method are found, on average, larger by a factor ranging between 1.3 and 2.8 than those obtained from the virial and projected mass estimators applied to the dominant galaxies of the groups. The most straightforward explanation of this fact is that there exists a fraction of missing mass within groups so far undetected, which can be ascribed to their dark haloes. Two anomalous results concerning UZC 268 and 340 represent cases involving the nature and the physical properties of the test groups. They were classified as "outliers" since their configurations did not match the fundamental requirement that a test group should be spherically symmetric and dynamically relaxed. From a statistical point of view, we attempt quantitative conclusions only for those groups that match the above requirements, even if our measures are not sufficiently good to provide small errors. The use of better and complete kinematic datasets and the application of more sophisticated methods of reconstruction of the external "true" mass distributions may largely improve our results and the level of statistical significance.

Finally, the efficiency demonstrated by our method in amplifying discrepant results may be useful to test whether a galaxy group is in a collapsing pre-virialized state or in a relaxed and symmetric configuration. Note also that this procedure could be applied to test the reliability of mass measurements obtained from other independent methods (for example, based on high-resolution X-ray images or gravitational lensing effect).

# TABLE 1

| UZC # | G $N_g$ | COORD. J2000 RA | DEC | V Km/s | $\sigma_V$ Km/s | $R_v$ Mpc/h | $M_v$ $10^{13}$ $M_{sun}/h$ | $F_{tidal}$ $10^{13}$ $M_{sun}h/Mpc^2$ | $R_t$ Mpc/h | $R_v/R_t$ | $\mu$ |
|---|---|---|---|---|---|---|---|---|---|---|---|
| 268 | 10 | 9 56 27.4 | +32 58 59 | 1454 | 57 | 1.14 | 0.26 | 3.6±0.3 | 0.27±0.02 | 4.24 | 18.0±10.8 |
| 291 | 8 | 10 17 20.4 | +21 54 18 | 1301 | 145 | 0.45 | 0.66 | 5.4±1.0 | 0.35±0.06 | 1.28 | 1.7±1.3 |
| 323 | 8 | 10 47 42.7 | +13 2 28 | 750 | 116 | 0.46 | 0.43 | 3.3±0.2 | 0.36±0.03 | 1.28 | 1.6±0.9 |
| 331 | 9 | 10 51 13.2 | +33 31 33 | 1648 | 135 | 0.62 | 0.79 | 2.4±0.4 | 0.54±0.09 | 1.15 | 1.3±0.9 |
| 340 | 17 | 10 58 32.9 | +17 42 16 | 1117 | 81 | 0.80 | 0.36 | 5.6±0.3 | 0.25±0.06 | 3.14 | 9.9±8.7 |
| 376 | 16 | 11 20 31.4 | +17 42 2 | 1110 | 240 | 0.60 | 2.45 | 5.6±0.5 | 0.66±0.18 | 0.90 | 0.8±0.9 |
| 466 | 9 | 12 11 35.3 | +13 29 12 | 2165 | 267 | 0.84 | 4.17 | 3.0±0.3 | 1.17±0.59 | 0.71 | 0.5±0.8 |
| 478 | 24 | 12 17 50.6 | +29 50 8 | 816 | 188 | 0.91 | 2.24 | 7.5±0.4 | 0.54±0.11 | 1.67 | 2.8±2.2 |
| 480 | 47 | 12 19 48.2 | +42 49 11 | 572 | 234 | 1.10 | 4.17 | 7.4±0.3 | 0.75±0.21 | 1.46 | 2.1±2.0 |
| 518 | 8 | 13 12 7.0 | +35 56 23 | 920 | 116 | 0.49 | 0.46 | 4.0±0.4 | 0.34±0.04 | 1.44 | 2.1±1.3 |
| 578 | 22 | 13 52 37.9 | +40 13 49 | 2381 | 181 | 0.75 | 1.74 | 1.6±0.2 | 1.04±0.22 | 0.71 | 0.5±0.4 |

**Fig. 1.** The cumulative amplitudes of $F$ as a function of the distance in 2 $h^{-1}Mpc$ bins from the barycenter of each UZC group. All curves are convergent within the limit of 20 $h^{-1}Mpc$. The outlier UZC 268 and 340 do not appear.

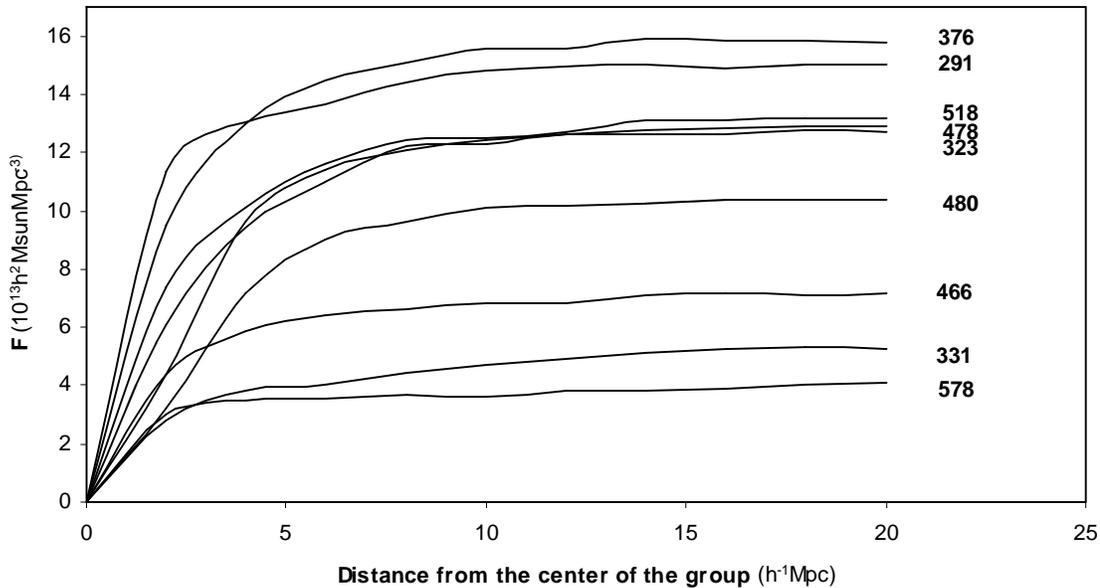

**Fig. 2.** For each UZC group, the value of $\mu_{PM}$ versus $\mu$. The outlier UZC 268 and 340 do not appear

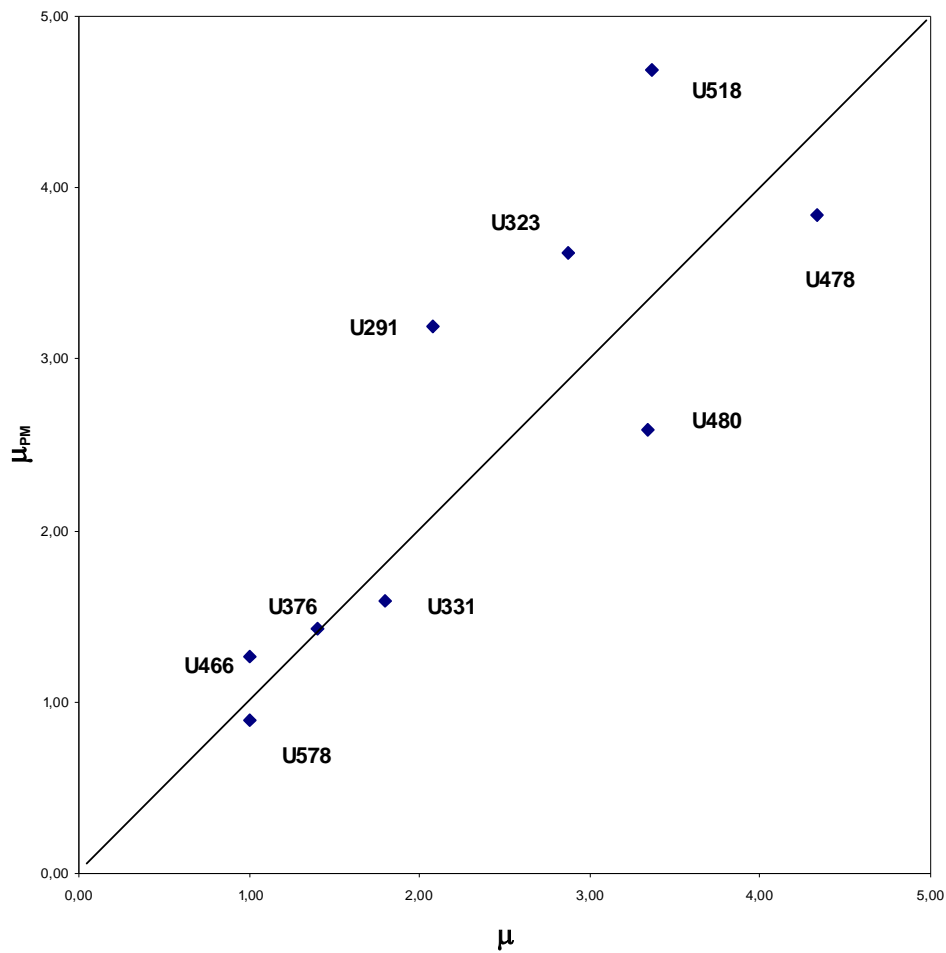